\documentclass{aa}
\usepackage{natbib}
\usepackage{graphicx}

\def\be{\begin{equation}}
\def\ee{\end{equation}}

\begin{document}

\title{The effect of HII regions on rotation measure of pulsars}

 \author{Dipanjan Mitra
        \inst{1}
        \and
        Richard Wielebinski
        \inst{1}
        \and
        Michael Kramer
        \inst{2}
        \and
        Axel Jessner
        \inst{1}}

\institute{ Max-Planck Institute f\"{u}r
Radioastronomie, Auf dem H\"ugel 69, D-53121, Bonn, Germany
\and University of Manchester, Jodrell Bank Observatory, Macclesfield SK11 9DL, UK}


\authorrunning{Mitra et al. }

\titlerunning{The effect of HII regions on rotation measure of pulsars }

\date{Received / Accepted }

\abstract{ 
 We have obtained new rotation measures for 11 pulsars 
observed with the Effelsberg 100-m radio telescope, in the direction
of the Perseus arm. Using a combination of 34 published 
and the 11 newly measured pulsar rotation measures we study the 
magnetic field structure towards the Perseus arm.
We find that two pulsars towards l$\sim$ 149$^{\circ}$ (Region 1)
and four pulsars towards l$\sim$113$^{\circ}$ (Region 2) lie behind
HII regions which seriously affect the pulsar rotation measures.
The rotation measure of PSR J2337+6151 seem to be affected by
its passage through the supernova remnant G114.3+0.3.
For Region 1, we are able to constrain 
the random component of the magnetic field to $5.7\mu$G. For the 
large-scale component of 
the Galactic magnetic field we determine a field strength of 
$1.7\pm1.0\mu$G. This average field is constant
on Galactic scales lying within the Galactic longitude range
of $85^{\circ} <$ l $ < 240^{\circ}$ and we find no evidence for 
large scale field reversal upto 5-6 kpc. We highlight
the great importance to include the effects of foreground emission
in any systematic study.}
\maketitle
\section{Introduction}
\label{sec1}

Pulsars are excellent probes of the Galactic interstellar medium (ISM).
Pulsar signals arrive at a radio telescope by passing through a complex medium
made up of gas, dust and magnetoionic plasma.  Already in the discovery paper
Hewish et al.~(1968) pointed out that the pulsed signals were delayed to lower
frequencies, typical of a dispersion in ionized hydrogen. The dispersion
measure (DM) of pulsars has been used, in connection with models of the
distribution of the column density of electrons, as a method of determining
the pulsar distance. In addition to dispersion, the linearly polarized pulsar
emission suffers Faraday rotation due to the electron density $n_e$ and a
component of Galactic magnetic field $B_{\parallel}$ along the propagation
path (e.g. see Lyne \& Smith 1989).  The combination of DM and rotation
measure (RM) can be used for determination of the local average magnetic field
(Smith 1968).

The dispersion of a pulsar signal is caused by the presence of free electrons
in the ISM. Several papers investigated the contributions to measured pulsar
DMs by HII regions (e.g.~Prentice \& ter Haar~1969) or by OB-stars (Grewing \&
Walmsley 1971). Complications arise from the observation that for HII regions
$n_e$ can vary significantly (e.g. Mitra \& Ramachandran 2001). These
observed variations in $n_e$ should lead to seemingly anomalous RMs for
pulsars and extragalactic (EG) sources, complicating the study of the Galactic
magnetic field.

Several pulsars are known to be associated with supernova remnants which can
also contribute to the measured RMs.  Studies of individual supernova remnants
(e.g. Vall\'ee \& Bignell 1983; Downes et al. 1981; Kim et al. 1988,
Uyan{\i}ker et al. 2001) suggest that RM values of $-150<$ RM $<$ 150
$\rm{rad~m^{-2}}$ are found in the Galactic plane \footnote{ Henceforth
wherever a numerical value of RM is quoted the unit used will be
$\rm{rad~m^{-2}}$.}.  In the supernova remnant G127.1+0.5 the RM is seen to
vary between $-30$ to $-130$ across the remnant (F\"{u}rst et al. 1984).  The
physical mechanism for the origin of these high RM's in supernova remnants is
unknown.  Kim et al.~(1988) studied point sources seen through G166.2+2.5 and
came up with values of RM $\sim 120 \pm 30$. Such complicated RM behaviour of
the supernova remnants clearly suggests that RM's pulsars will have
significant contributions due to its passage through the supernova remnants.

The observed RMs of point-like EG sources vary significantly.  Early
observations of Centaurus A by Cooper \& Price (1962) showed Faraday rotation
with a RM of $-70$ that was attributed to radio emission passing through the
Galactic halo.  Studying the RM of EG point sources became `big business',
culminating in the work by Simard-Normandin \& Kronberg (1980) where data of
552 sources were presented. The observed RM values varied by about $\sim\pm$
300 (see also the catalogue of Broten et al. 1988), but in most cases high RM
values are probably intrinsic to the source.  Indeed, only a very few of the
previously catalogued sources are really seen through the plane of the Galaxy.
The sample of Simard-Normandin \& Kronberg (1980), for instance, was
restricted to sources {\em outside} the band of Galactic latitude $b \le \pm
5^{\circ}$.  Recent observations of a dense sample of sources {\em within}
$|b| \le 4^{\circ}$ by Brown \& Taylor (2001), however, resulted in RM values
of $\sim\pm$ 400, clearly showing that the RM of sources seen through the
plane of the Galaxy is much higher than for sources distributed over the rest
of the sky. In the same study, Brown \& Taylor (2001) identified regions
in the outer Galaxy, where the RMs of EG sources are seen to change
anomalously due to the radio emission's passage through complex emission
structure seen in the total intensity radio continuum emission at 21 cm.
They interpret these anomalies resulting from local reversals of magnetic
fields. 

From all the above observations we must indeed expect that the measured RM
of pulsars and EG sources will be seriously affected by the passage through
local features of the ISM.  The effect is particularly important for pulsars
as the majority of them lie in the Galactic plane.  HII regions associated
with local fluctuations in the Galactic plane complicate
the interpretation of data when trying to disentangle the contributions from
Galactic and intrinsic origin.

Rand \& Kulkarni (1989) have discussed the
effect of large scale HII regions and note that a region of enhanced
turbulence at about $45^{\circ} < l < 75^{\circ}$, $10^{\circ} < b <
65^{\circ}$. This may be associated with the
North Polar Spur, which affects
the pulsar RMs significantly.  Such effects however
have not been taken into account for all lines-of-sights to pulsars
systematically when trying to
study the large-scale regular component of the magnetic field in our
Galaxy (e.g. Rand \& Kulkarni 1989, Rand \& Lyne 1994, Indrani \&
Deshpande 1998, Han et al. 1999, Frick et al.~2001). 
The most careful analysis was that of Indrani \& Deshpande (1998)
as they rejected all pulsars for which the estimated $B_{\parallel}$ was
higher than 3$\mu$G an issue which we will address later.
In an attempt to ascertain
how foreground HII regions affect the RM of pulsars, in this paper we
concentrate on a small section of the Galaxy namely towards the
Perseus arm, and investigate the effects of small scale HII regions on
pulsar RMs (Section~\ref{sec3}). In order to increase the number of
pulsars with known RM, we obtained new RMs of pulsars using the
Effelsberg radio telescope as described in Section~\ref{sec2} .  With
the increased statistics and a carefully selected set of pulsars we
study the regular component of the magnetic field as described in
Section~\ref{sec4}. Finally, in Section~\ref{sec5} we discuss the
implication of our findings.

\section{Pulsar rotation measures using the Effelsberg Radio Telescope}
\label{sec2}

\begin{figure*}
\begin{tabular}{@{}lr@{}}
{\mbox{\includegraphics[height=8cm, width=5cm, angle=-90]{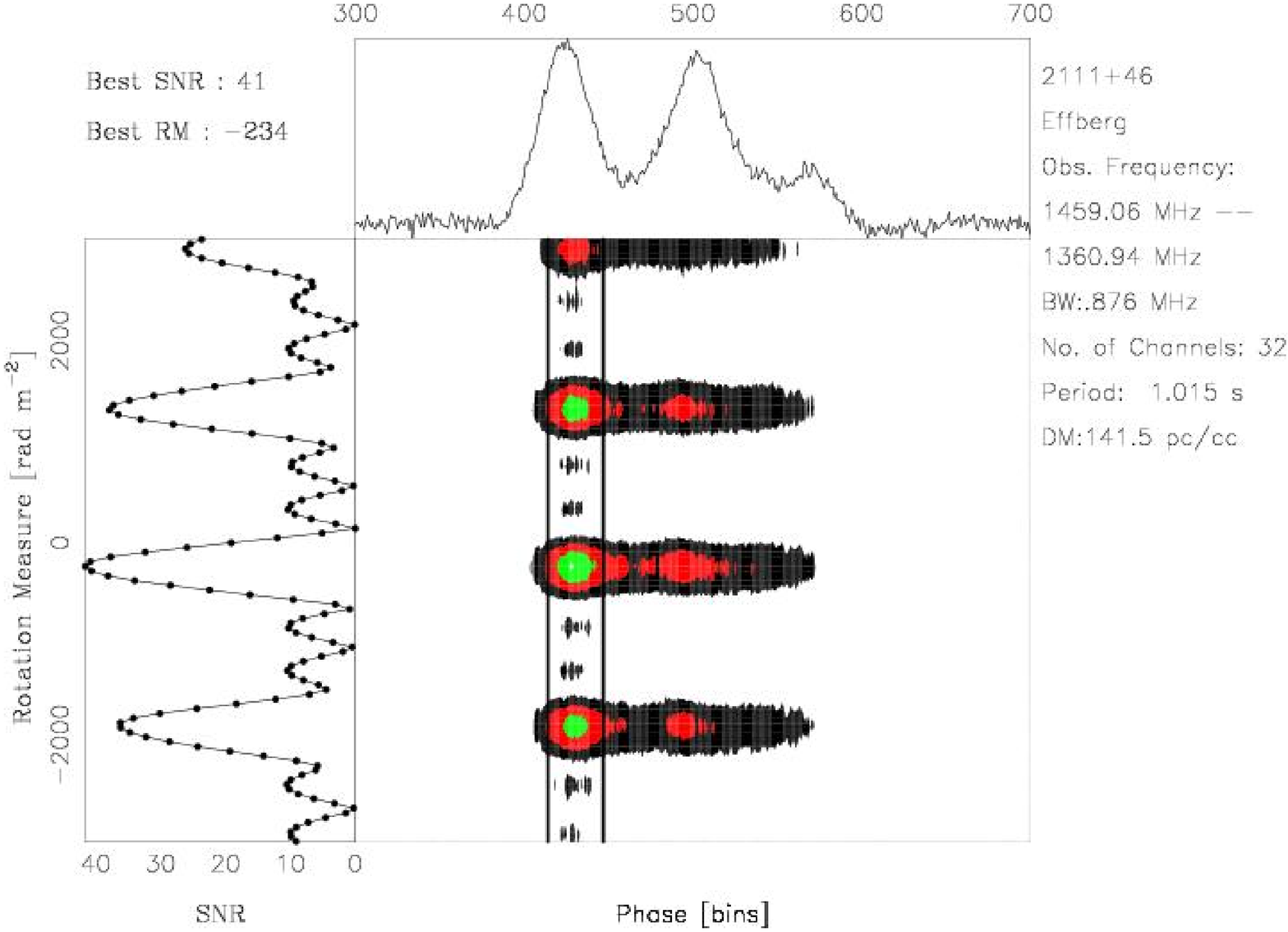}}}&
{\mbox{\includegraphics[height=8cm, width=5cm, angle=-90]{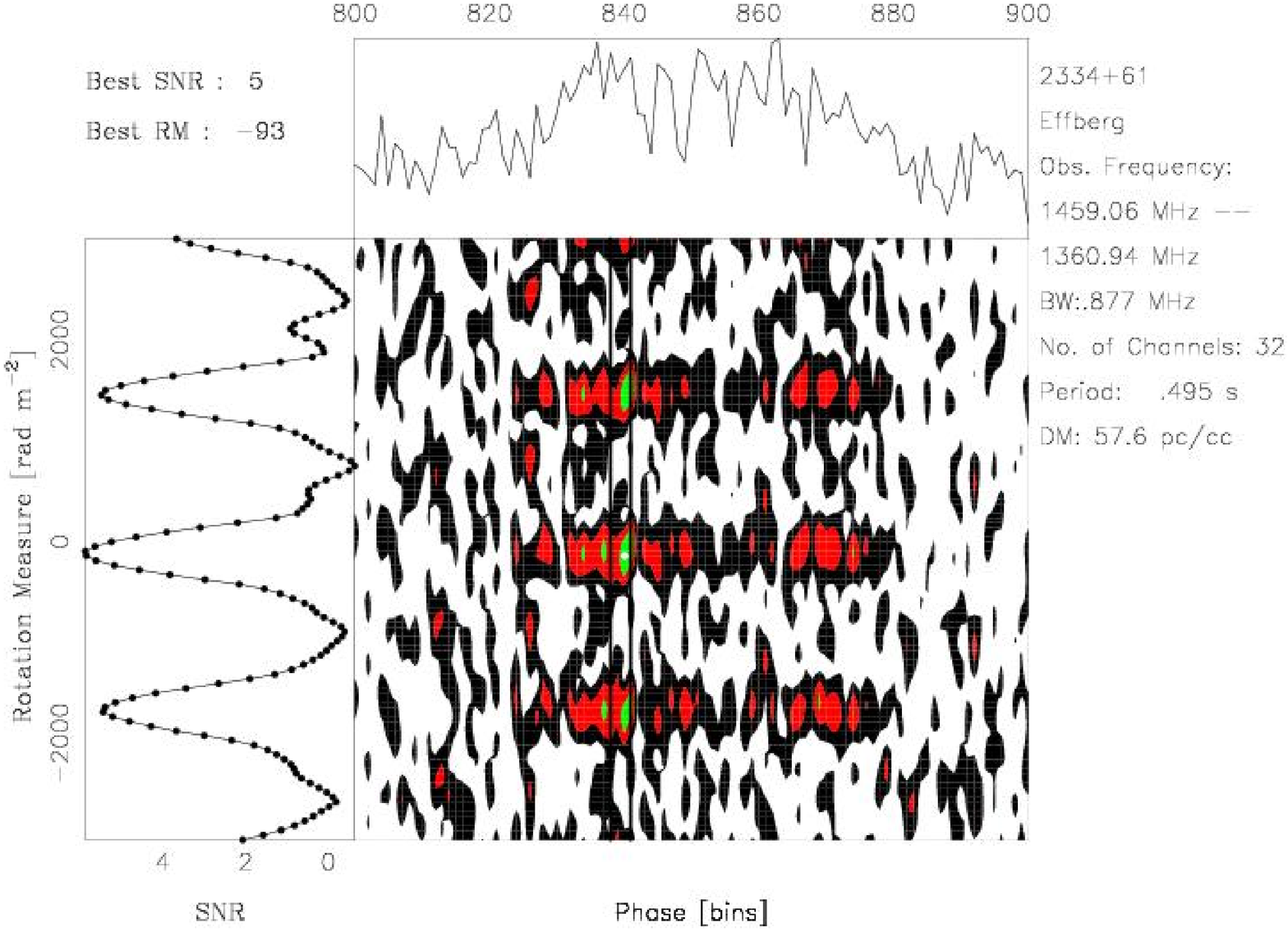}}}\\
\end{tabular}
\caption[]{Illustration of the method used to determine pulsar RMs showing
examples of PSRs J2113+4644 (left) and  J2337+6151 (right). 
The top panel in each plot shows the total intensity of the pulsar. 
The central panel shows a contour plot of
the S/N of the linear polarization as a function of assumed RM values
(vertical axis) and pulse phase (horizontal axis). Pulse phases within
the shown vertical lines have been collapsed to obtain the S/N as a
function of RM that is displayed in the left panel.}

\label{fig1}
\end{figure*}

Currently, about 1500 pulsars are known. While RMs are
published only for about 350 of them, most of the major projects over
the last decade were undertaken in the southern hemisphere (e.g. Han
et al. 1999). Thus there remain a large number of pulsars for which
RM are largely unknown.  In order to augment the
sample, in particular towards our selected region of the Perseus
arm, we have undertaken a project of measuring RMs of pulsars using
the 100-m Effelsberg radio telescope.

A sample of pulsars in the northern hemisphere ($\delta$ $>0^{\circ}$) 
for which no RM are available, was
chosen from the Taylor, Manchester \& Lyne (1993, updated version
1995) pulsar catalogue.  Observations were made using the
Effelsberg telescope in October/November 2001 using the 1.4 GHz
HEMT receiver installed at the prime focus of the telescope. This
receiver is tunable between 1.3 and 1.7 GHz. For median elevations the
receiver has a noise temperature of 25 K. Dual circularly polarized
signals, LHC and RHC, are mixed down to an intermediate
frequency of 150 MHz and fed into the coherently de-dispersing
Effelsberg Berkeley Pulsar Processor (EBPP, for technical details
see Backer et al. 1997). Within the EBPP, the signal is split
into four sub-bands using four down-converters. Each down-converter
consists of two mixer filter modules (MFs) and one local oscillator
(LO).  Each MF receives two orthogonally polarized IF signals that are
fed by the common LO, before they are passed into digital filter
boards (DFBs). Each DFB provides a further sub-division of the signals
into eight channels, which are then coherently de-dispersed in
hardware by de-disperser boards (DBs). At the end a total of 32
channels for each polarization are obtained.  Depending on the
frequencies of the four LO's the sub-bands can be spread across the
available observable receiver bandpass 
in four sub-bands, symmetrically placed with
respect to the centre frequency of observation.  This option of the
EBPP is quite useful for better RM estimates at 1.4 GHz where
the sub-bands can be spread to obtain total bandpass of $\sim$100 MHz.
The
bandwidth of each final frequency channels depends on the DM of the
observed pulsar, but is limited to 0.875 MHz for the polarization mode
at around 1.4 GHz. In total, a bandwidth of 28 MHz is available,
spread around in four sub-bands of 7 MHz.  All signals are detected and
folded in phase with the topocentric pulsar period.  Each
sub-integration which typically lasts for 180--300 s is transferred to
a computer for offline processing.

Each observation of a pulsar to measure its RM is preceded
by a calibration scan where the signal of a noise diode is injected
into the waveguide following the feed horn. Pulsar and calibration
scans are used to construct the pulsar's Stokes parameters $I$,
$Q$, $U$ and $V$ for each frequency channel.  Parallactic angle
corrections are applied to Stokes $U$ and $Q$ before all
sub-integrations for a given pulsar are finally aligned in time and
added to form the integrated Stokes parameters for all 32 frequency
channels.

A search in rotation measure is performed by maximizing the
linearly polarized intensity across the pulse phase.  Stokes $U$ and
$Q$ are appropriately rotated for a given value of the RM and then
added for all the frequency channels to produce the linearly
polarized intensity $ L = \sqrt{U^2+Q^2}$.  Phase bins across the
pulse where the signal-to-noise ratio (S/N) of $L$ exceeds a threshold
of 3 are collapsed to find the average S/N as a function of RM. The
curve's most significant peak with a S/N $>$5 is identified as being
close to the pulsar's RM.  A finer search is then performed close to
this peak to improve the RM estimate.

Since position angles are ambiguous by 180$^{\circ}$, searches are performed
for an increased RM range until a second peak in the S/N vs RM curve
is obtained. The solution with the highest S/N can be clearly
identified and is taken as the measured pulsar RM. Figure 1 gives two
examples of the described technique for PSRs J2113+4644 and PSR
J2337+6151.  Although the S/N is very different, the RM can be
accurately determined in both cases.

Uncertainties in the measured RMs are estimated directly from the S/N
vs RM curves. The curves show a rather symmetric behaviour around the
peak and positive and negative errors are determined by reading off
the RM values corresponding to those values where the S/N decreases by
one unit. Since 32 independent channels are used to produce these S/N
curves, the resultant error is divided by $\sqrt{32}$. One way to
check the validity of our method is to choose pulsars where in a given
phase bin the S/N of $L$ of all the available frequency channels is
significantly high ($>$ 10). In this case we can construct the
position angle $\Phi = 0.5 {\rm tan}^{-1}(U/Q)$ and fit a straight
line of the form $\Phi = \Phi_{\circ} + {\rm RM}~\lambda^2$, where
$\lambda$ is the wavelength in meters. The resulting RMs and their
formal errors obtained by this method are in good agreement with the
procedure described above.  For PSR J2113+4644 shown in
Fig.~\ref{fig1}, for instance, both methods yield a RM of 230$\pm$8, 
which is also in good
agreement with the value available from the literature of 224$\pm$2.
Other test pulsars with known rotation measures yield a similar good
agreement with the earlier measurements. 

We concentrated our measurements on pulsars located in a region within
$85^{\circ} < l <245^{\circ}$ and $-10^{\circ} < b <10^{\circ}$.
There are 56 known pulsars in this direction. Table~\ref{tab1} lists
44 of them for which RMs have been determined. Those pulsars with RM
measurements obtained in this paper are highlighted by boldface in the table.

\begin{figure*}
\resizebox{\hsize}{!}
{\includegraphics[angle=-90 ]{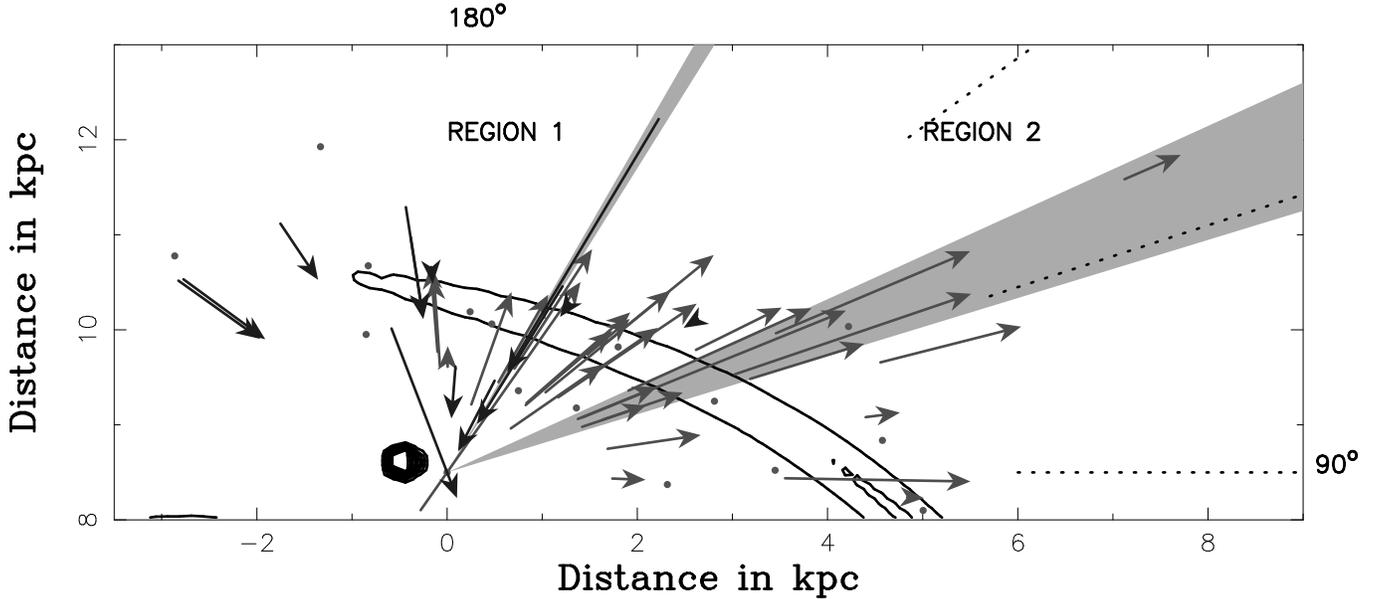}}
\caption[]{Distribution of pulsars in the Galaxy towards the
Perseus arm within a Galactic latitude range of $-10^{\circ} < b <10^{\circ}$. 
The Sun is located at ordinate 8.5 kpc and abscissa 0 kpc. Pulsars with
arrows directed towards the observer have positive RM and 
those away from the observer have negative RM. The size
of the arrows are proportional to the $B_{\parallel}$ as calculated using 
Eqn.~\ref{eq1} with the pulsars located at the center of the arrows. 
The dots are pulsars for which RM values are not available.}
\label{fig2}
\end{figure*}

\begin{figure*}
\resizebox{\hsize}{!}
{\includegraphics[height=5cm,angle=-90]{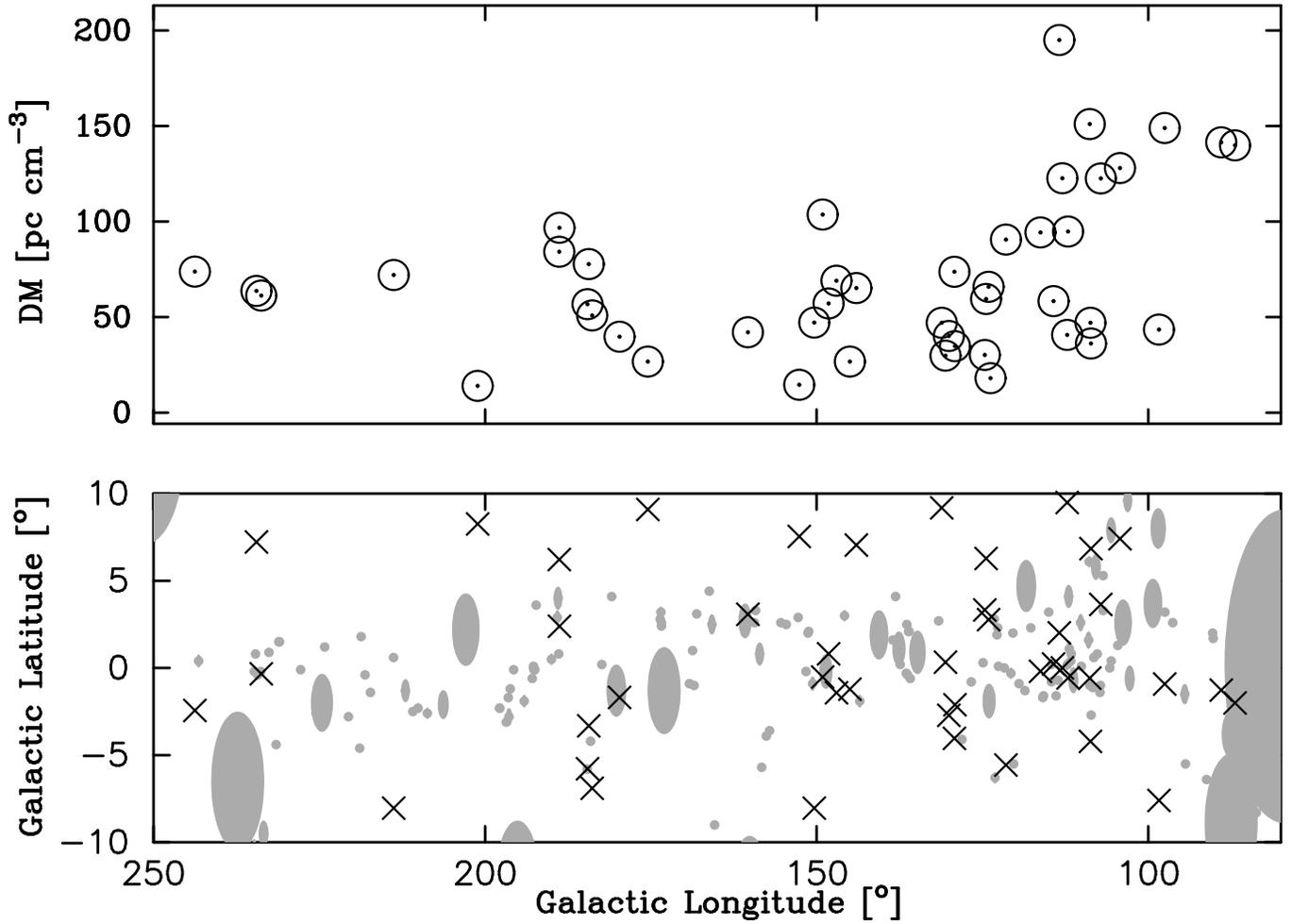}}
\caption[]{{\em (Bottom)} Distribution of HII regions as obtained from
the Sharpless (1959) catalogue along the Galactic plane with
longitudes $85^{\circ} < l <245^{\circ}$ and Galactic latitudes
$-10^{\circ} < b <10^{\circ}$.  Approximate sizes are depicted as gray ellipses
using the angular extent as given by the catalogue. Note that 
detailed size and shape of the HII regions can be different from 
what is shown here. The positions of
pulsars for which RM information is available are overlayed on the
plot with crosses.
{\em (Top)} Corresponding DMs observed for the pulsars shown.}
\label{fig3}
\end{figure*}

\begin{figure*}
\begin{tabular}{@{}lr@{}}
{\mbox{\includegraphics[height=6cm, width=6cm]{region1_halpha.epsi}}}&
{\mbox{\includegraphics[height=6cm, width=6cm]{region2_halpha.epsi}}}\\
{\mbox{\includegraphics[width=6cm,height=6cm]{region1_21cm.epsi}}}&
{\mbox{\includegraphics[width=6cm,height=6cm]{region2_21cm.epsi}}}\\
{\mbox{\includegraphics[width=6cm,height=6cm]{region1_11cm.epsi}}}&
{\mbox{\includegraphics[width=6cm,height=6cm]{region2_11cm.epsi}}}\\
\end{tabular}
\caption[]{Pulsar positions overlayed on {\em (left)} Region 1
and {\em (right)} Region 2 
as seen in several frequency ranges. The upper most
panels shows an H$\alpha$ map, while the middle and bottom plots
show radio continuum maps at $\lambda$21cm and $\lambda$11cm,
respectively. The crosses in the figures indicates the position of 
the pulsars. The (RM,DM) and the pulsar name are indicated next to the crosses
in the topmost figures of the right and left panel.}
\label{fig4}
\end{figure*}

\section{The Perseus Arm }
\label{sec3}

Our aim is to study the effect of highly ionized HII regions on pulsar RMs.
Noting that the distribution of HII regions peaks within a few degrees of the
Galactic plane, we restrict the pulsars studied to $|b| < 10^{\circ}$,
i.e.~ensuring that all pulsars lie within a height, $z$, above the Galactic
plane of 1 kpc (see Table~\ref{tab1}), which is also the typical electron
density scale height.  In Figure~\ref{fig2} we plot the top view of the Galaxy
as given by the spiral arm model of Georgelin \& Georgelin (1976) along
with the distribution of pulsars\footnote{for the latest catalogue information
  visit {\tt http://www.atnf.csiro.au/research/pulsar/catalogue}} that are
located within our selected latitude range.  Pulsar distances are estimated
from the Cordes \& Lazio (2002, CL02 hereafter) electron density model.  The
CL02 model is a significant improvement over the Taylor \& Cordes (1993)
electron density model as it takes into account the clumped structure of the
ISM thus giving improved distance estimates to pulsars.  Note that we also
include PSR J0056+4756 in our study. With $b=-14^{\circ}$ it lies outside our
nominal $|b|<10^\circ$ strip, but due to its small estimated distance, it has
a small $z$-height of only $-0.26$ kpc.

The RM and DM of pulsars is used to estimate the average parallel
component of the regular magnetic field as,
\begin{equation}
B_{\parallel} = 1.232~ {\rm (RM/DM)}~~~\mu {\rm G } 
\label{eq1}
\end{equation}
where RM is in rad m$^{-2}$ and DM in pc cm$^{-3}$. 
This estimate holds only if the variations of 
$B_{\parallel}$ and $n_e$ are 
uncorrelated along the line of sight (Beck 2001).
As evident from Figure~\ref{fig2}, near the solar neighbourhood
there is a regular component of the magnetic field which changes
sign due to the line of sight effect (Rand \& Lyne 1994, Indrani \& Deshpande 
1997, Han et al. 1999, Frick et al. 2001). 

In the bottom panel of Fig.~\ref{fig3} we plot the distribution of  
HII regions from the Sharpless (1959) catalogue that are 
located within $85^{\circ} < l < 245^{\circ}$ and  $|b| < 10^{\circ}$, 
and along with the pulsars for which RMs are available.
Among the 45 pulsars with measured RM, 14 of them have positive RMs.
Around $l \sim 149^{\circ}$ there are two pulsars with high 
positive RM (cf.~Fig.~\ref{fig2})
which we refer to as Region 1 hereafter (see also Table~\ref{tab1}). 
If the regular component of the 
magnetic field follows the Perseus arm, and the pulsars 
sample only the regular component of the magnetic field, then
all the pulsars in this longitude range should have negative 
RMs. Thus the positive RMs could arise either due to reversal
of the regular component of the magnetic field or due to some
locally turbulent component of the field which disturbs the regular 
field sufficiently. Such turbulent components
could also be responsible for anomalously high negative
RMs.

In the top panel of Fig.~\ref{fig3}, we plot the DM of those
pulsars shown in the bottom panel. Apart from Region 1, a
increase in both RM and DM of pulsars is seen around Galactic
longitude of $l\sim 113^{\circ}$, which we call Region 2 hereafter
(see also Tab.~\ref{tab1}).  Pulsars in Region 1 and Region 2 have
lines-of-sight either directly through or close to HII regions. RM of
pulsars behind these regions are thus susceptable to be affected
seriously by these HII regions.  We investigate this issue further
below.
\begin{figure}
{\includegraphics[height=7cm, width=8cm]{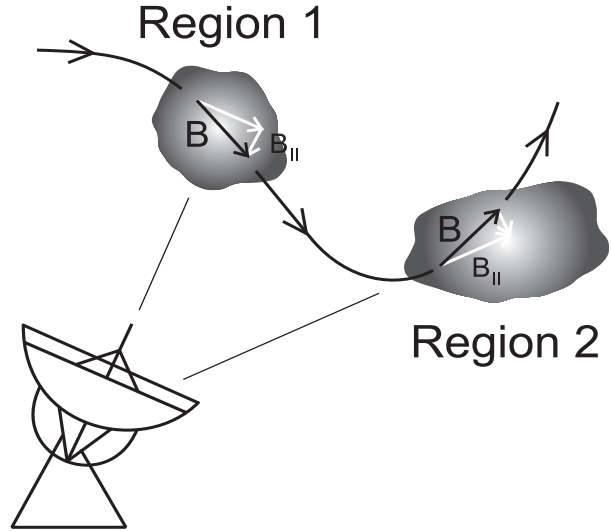}}
\caption[]{Schematic model of the magnetic
field in the direction of the Perseus arm. See text for details.} 
\label{schem}
\end{figure}

\begin{figure}
{\includegraphics[height=8cm, width=5cm, angle=-90]{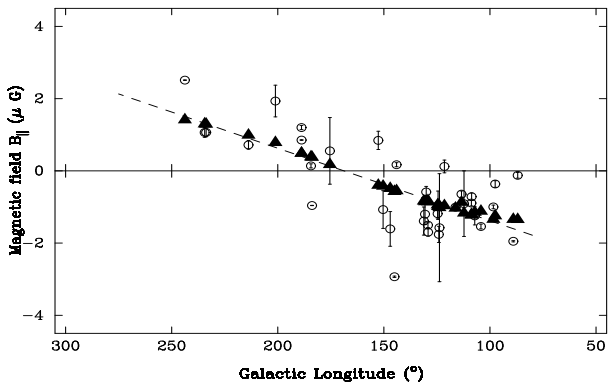}}
\caption[]{Variation of the estimated $B_{\parallel}$
with Galactic longitude. The dashed line is a linear fit to 
the data and the solid triangles are the fit by using the model  
described in section.~\ref{sec4}. The open circles 
are the observed $B_{\parallel}$ for the selected pulsars.}
\label{fig5}
\end{figure}

\begin{figure*}
\begin{tabular}{@{}lr@{}}
{\mbox{\includegraphics[height=6cm, width=6cm,angle=-90]{sec1.ps}}}&
{\mbox{\includegraphics[height=6cm, width=6cm,angle=-90]{sec1dist.ps}}}\\
{\mbox{\includegraphics[width=6cm,height=6cm,angle=-90]{sec2.ps}}}&
{\mbox{\includegraphics[width=6cm,height=6cm,angle=-90]{sec2dist.ps}}}\\
{\mbox{\includegraphics[width=6cm,height=6cm,angle=-90]{sec3.ps}}}&
{\mbox{\includegraphics[width=6cm,height=6cm,angle=-90]{sec3dist.ps}}}\\
\end{tabular}
\caption[]{{\em Left:}Variation of RM versus DM of pulsars lying within
three Galactic longitude range are shown. The open circles are pulsars
which lie behind HII regions and supernova remnants. The filled circles 
are pulsars which are apparently not affected by HII regions along the line
of sight. The dashed line is the straight line fit to data points. The 
$B_{\parallel}$ obtained from the fits are shown in the figure.
{\em Right:} Variation of RM with distances are shown for the three
different sections. The symbols are same as that in the left panels. See 
text for further details. } 
\label{fig6}
\end{figure*}

\begin{figure}
{\includegraphics[height=8cm, width=5cm, angle=-90]{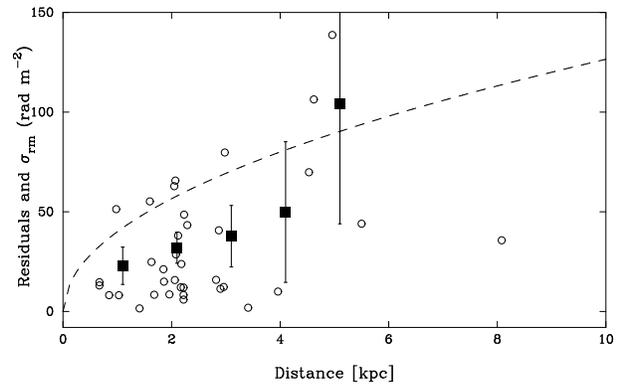}}
\caption[]{In the above plot the residuals of RM's and $\sigma_{\rm rm}$ 
are plotted as a function of distance to the pulsar. 
The residuals of RM 
are denoted by open circles, and the estimated $\sigma_{\rm rm}$ are denoted
by filled squares. The dashed line corresponds to $40 \sqrt{D}$.
See text for further detail.}
\label{fig7}
\end{figure}


\subsection{ Region 1: $l = 149^{\circ}$}

In this region data are available for four pulsars and three
extra-galactic sources. The pulsars J0357+5236 and J0358+5413
have positive RMs while the nearby objects PSRs J0332+5434 and
J0343+5312 have negative RMs. Three EG sources in this
area (Kronberg, private communication) also have negative
RMs. The DM of PSR J0357+5236 is particularly high, much higher than
for all other nearby pulsars. We have overlaid the position of the
four pulsars on a H$\alpha$ image taken from the Virginia Tech
Spectral-Line Survey (VTSS, Simonetti et al. 1996)  in Fig.~\ref{fig4}. It shows
that PSR J0357+5236 is seen directly towards a bright HII region,
explaining the high DM of the object. The change in sign pertains to
two pulsars, both seen through the bright HII region. The
DM of the other pulsars is similar suggesting them to be close to each
other. There are two more observables  
givng evidence of dense ionized HII region towards 
a pulsar. Firstly it is expected that electron density 
fluctuation in strongly ionized HII regions will result in
scatter broadening of pulsar signals (see Mitra \& Ramachandran 2001 and
references therein). Secondly the emission measures
towards these directions should be significantly higher.
While the pulse shape of PSR J0357+5236 shows an apparent signature
of scatter broadening in the form of an conventional exponential 
decay of the pulse profile, we are not aware of any measurements 
available for this pulsar in the literature. In fact no conclusion 
can be drawn regarding this without 
any detailed analysis as scattering measurements are influenced strongly
by the intrinsic pulse shape (L\"{o}hmer et. al. 2001). The 
scatter broadening time for J0358+5413 is 2.3$\times 10^{-4}$ msec, 
almost 4 times higher than
the nearby pulsar J0332+5434 (refer pulsar catalogue Taylor, Manchester \& Lyne, 1993). 
The emission measures towards PSR J0357+5236 and J0358+5413 
are about 103 and 29 pc cm$^{-6}$
compared to an average value of 10 pc cm$^{-6}$ towards the nearby objects
(data from M\"{u}ller et. al. 2003, private communication).
These evidences further reinforces the conclusion that PSR 
J0357+5236 and J0358+5413 are potentially behind the HII region S205.

 Overlaying the pulsar positions onto $\lambda$21cm and
$\lambda$11cm radio continuum maps (Reich et al. 1990, F\"{u}rst et al. 1990) in
Fig.~\ref{fig4}, we notice that the pulsars are seen through complex
Galactic emission regions. The fact that at $\lambda$11cm 
the diffuse emission is less than at $\lambda$21cm suggests 
the presence of strong non-thermal emission in this direction.

A large region near $l\sim145^{\circ}$ is occupied by the Camelopardalis
dark clouds which are amongst the closest dust formations in the solar 
vicinity (Zdanavicius et al. 2001). The HII region S205 (Sharpless 1959)
which is of interest
has its northern half in the Camelopardalis and the southern half in 
the Perseus constellation. S205 is thought to be ionized by the O8 star 
HD 24451 and is also listed in the catalogue of bright nebula by Lynds (1965).
Using the CO velocities of the optical HII emission, 
Fich \& Blitz (1984) quote
a distance for S205 of 900$\pm$300 pc. Using the angular diameter 
of the optical HII region, they also obtain a linear size of 31.4$\pm$10.5 pc. 

Having identified PSRs J0357+5236 and J0358+5413 to be lying behind S205,
these pulsars can now be used to put constrains on $n_e$ and $B_{\parallel}$
for this HII region. Apart from these two pulsars the other objects in the
nearby region have an average DM of $\sim$ 50 pc cm$^{-3}$. If we attribute
the increased DM of 103.65 pc cm$^{-3}$ for PSR J0357+5236 entirely to S205,
then the excess DM is $\sim$53 pc cm$^{-3}$. Further we can use S205's
transversal scale of 30 pc to estimate $n_e$ to be 1.8 cm$^{-3}$. The average
RM of the nearby pulsars is $\sim -50$. With the estimated $n_e$ = 1.8
cm$^{-3}$ over a distance of 30 pc, we infer $B_{\parallel}\sim5.7\mu$G which
is sufficient to explain PSR J0357+5236's increased RM of 261. To explain the
lower RM and DM values of PSR J0358+5413 a nominal value of only $n_e\sim
0.75$ cm$^{-3}$ would be sufficient while keeping $B_{\parallel}$ constant.
Of course, apart from this simple model,
another way to explain the increased DM and RM values would be to
incorporate variations in both $n_e$ and $B_{\parallel}$. This however would
require the HII region to modulate the magnetic field within itself which is
of the order of several $\mu$G. Albeit such strong magnetic fields might not
be easy to generate in HII regions.  Thus it is possible that S205 is
associated with a local loop of the magnetic field where the field twists in
the direction of the observer while the magnetic field strength across the 30
pc scale estimated for S205 is rather constant.

\subsection{ Region 2: $l = 113^{\circ}$}
 
The other region of RM variation is seen around $l\sim
113^{\circ}$. There are 7 pulsars in this direction which are
listed in Tab.~\ref{tab1}. In the right hand top panel of
Fig.~\ref{fig4} we have overlayed the pulsar position on the
H$\alpha$ map observed by the Wisconsin H$\alpha$ mapper (WHAM,
Haffner et al. 2001). Apparently, PSRs J2229+6205, J2308+5547, and
J2325+6313 are in regions free from enhanced emission.  The estimated
average $B_{\parallel}$ for these pulsars is about 1$\mu$G.  From the low
DM of 47 pc cm$^{-3}$ for J2308+5547 the estimated model distance is 2.2
kpc. The other two objects PSRs J2229+6205 and J2325+6313 have high
DMs of 122.6 and 195 pc cm$^{-3}$ which correspond to distances of
4 and 8 kpc, respectively. The nearly constant estimated
$B_{\parallel}$ for these two pulsars suggests that the magnitude of
the regular component of the magnetic field is constant over a large
distance beyond the Perseus arm (see Tab.~\ref{tab1}).

The pulsars J2257+5909, J2321+6024 and J2326+6113 are seen to lie
towards a complex of enhanced H$\alpha$ emission (Fig.~\ref{fig3} and
~\ref{fig4}).  The increase in DM for these objects is also associated
with an increase of RM giving an average $B_{\parallel}$ of
2.6$\mu$G. PSR J2337+6151 is known to be associated with the supernova
remnant G114.3+0.3 (Kulkarni et al. 1993, F\"{u}rst et al. 1993) and is
prone to have an anomalous RM behaviour.  In fact, the estimated
$B_{\parallel}$ of 2.1$\mu$G is higher than the regular component in 
the Perseus arm. This
also supports the fact that the pulsar is associated with the
supernova remnant.

The overlays of these pulsars on the radio continuum total intensity
maps of $\lambda$21cm and $\lambda$11cm 
are shown in Fig.~\ref{fig4} (Reich et. al. 1990, F\"{u}rst et. al. 1990).  
The fading of several structures between these two frequencies
shows the non-thermal behaviour of these regions while there are some
complexes which seem to remain thermal. Another possibility for
this apparent fading might that the emission is too faint to be
detected in usually less sensitive high-frequency radio surveys.  
Unlike Region 1, this area is far more
complex in terms of distribution of HII regions. 
paper.

For PSR J2257+5909 a number of HII regions may contribute to
the enhanced electron density along the path, i.e.~S152, 
BFS14 and/or BFS15 (Blitz et al. 1982).  
The distance estimated from CO velocities for S152, 
which is closest to the pulsar, is 3.6$\pm$1.1 kpc. 
The estimated CL02 model distance to the pulsar is 4.5 kpc,
thus placing the pulsar behind the HII regions. 

 PSRs J2321+6024 and J2326+6113 lie close to the HII regions S161A,
S161B, S162 and S163. S161A and S161B are considered to be two
HII regions almost superposed on each other with a distance of
2.8$\pm$0.9 kpc (Blitz et. al. 1982). With S162's distance
estimate of 3.5$\pm$1.1 kpc, all three regions are close to PSR
J2321+6024 whose distance using the CL02 model is $\sim$3
kpc. S163 has a distance of 2.3$\pm$0.7 kpc and is
closer to PSR J2326+6113 whose model distance is 4.5 kpc, again
indicating that the pulsar is behind the HII region. Note that
within the uncertainties all HII regions are located at a distance of
approximately 3 kpc. Again, it seems likely that the HII complexes are
associated with a local loop of the magnetic field, but now the
direction of the field is directed away from the observer. The
situation of both Region 1 and Region 2 is illustrated schematically
in Fig.~\ref{schem}.

It should be noted that although while estimating distances to pulsars the
CL02 model tries to take into account the enhanced electron density due to HII
regions for specific lines of sight these distances might still be inaccurate.
However, without complete knowledge about the length scales and the $n_e$ of
these HII regions, it is not possible to calculate these distances accurately.

\section{The regular component of the magnetic field}
\label{sec4}

Our investigation has concentrated on a region of $85^{\circ} < l
< 245^{\circ}$ and $|b|<10^{\circ}$.  In order to study the large
scale configuration of the regular Galactic magnetic field towards the
Perseus arm, we now consider all pulsars in this region, summarized in
Table~\ref{tab1}. Having shown that anomalous variations of RM and DM
can be  associated with HII regions, in particular
in our so-called Regions 1 and 2, we are now in the position to use
Table~\ref{tab1} to construct a sample of sources that is free from
such misleading effects.

The sample listed in Table \ref{tab1} includes a few pulsars which are
believed to be associated with supernova remnants. PSRs J0502+4654 and
PSR J0538+2817 are located within the boundaries of the supernova
remnants G160.9+2.6 (HB9) and G180.0$-$1.7 (S147), respectively. 
From surface
brightness-diameter relations (e.g. Milne 1979) the distance to
G160.9+2.6 is thought to be $\sim$1.2 kpc (Leahy \& Roger 1991) which
is in good agreement with the distance of $\sim$1.4 kpc to PSR
J0502+4654 estimated from the CL02 model, suggesting
a physical association of the pulsar with the supernova remnant. PSR
J0538+2817 has a distance of $\sim$1.2 kpc and is thought to be
physically associated with the G180.0$-$1.7 whose distance is
estimated to be $\sim$0.8 kpc as deduced from interstellar redenning
measurements ( Fesen et al. 1985, see also Anderson et
al. 1996) and $\sim$1.6 kpc based on surface-brightness diameter
relationship (Sofue et al. 1980). As pointed out before,
it is uncertain about how a supernova remnant affects the RM of
such pulsars, and thus we exclude these objects from further discussions.
For similar reasons, we do not consider the Crab pulsar, PSR J0534+220.

The remaining 36 pulsars form a carefully selected sample with 
RM and DM values that are free from any obvious anomalies
that might be introduced due to bright HII regions or supernova
remnants. For all these pulsars, labeled with a `G'
in Table.~\ref{tab1}, we use Eqn.~(1) to compute the weighted average
of $B_{\parallel}$, and we plot the resulting values 
as a function of Galactic longitude in Figure~\ref{fig5}.

Inspecting Fig.~\ref{fig5}, there is some tendency for 
the average $B_{\parallel}$ to
decrease from $l\sim 90^{\circ}$ to $l\sim 150^{\circ}$. Beyond $l\sim
150^{\circ}$, we notice an increase in mean B$_{\parallel}$ from
negative to positive values. In the simplest model, the distribution
of points in Fig.~\ref{fig5} could be described by fitting a linear
dependence to the data as shown by the dashed line. Since the
estimated uncertainties in the average
$B_{\parallel}$ are much smaller than the
spread in the data points, however, an unweighted least-squares fit
was performed to derive the shown linear dependence which is clearly only a
crude description of the data. 

In order to forge a better
understanding, we can instead try to test whether the data are result of the
magnetic field following the local Perseus arm.  The
motivation for this comes from examples of external galaxies, where
the large scale inter-arm field seems to trace the optical spiral arms 
with a slight phase offset (Beck 2001). We consider a model where the regular
uniform magnetic field of magnitude $B_{\rm o}$ traces the spiral arm
as depicted by Georgelin \& Georgelin (1976).  
In our model this field is meant to be located  not only on the spiral arm, but also in the
inter-arm regions between the solar system and the Perseus arm as well as beyond the arm. 
We use the procedure of cubic spline fit as adopted by Taylor \& Cordes 
(1993) to construct the spiral arm and the direction of the field lines. 
The angle between the field lines and the line of sight varies with distance and Galactic
longitude so that  varying amounts of  the global field will be projected onto the line of sight when
a radiowave travels from the pulsar to solar system. To account for that effect we
integrated  numerically over the projected B$_{\rm o}$ along the line of sight for every
source to obtain their effective $B_{\parallel}$. 
This dependence of $B_{\parallel}$ is then used to fit the data points.  
In addition we allow for another free parameter to be
determined, i.e.~$\theta$ which describes the possibility that the
so-called `magnetic arm' might actually be at a phase offset with respect to
the spiral arm, a circumstance which is also commonly observed in
external galaxies (Beck~2001). A positive value of $\theta$ 
means that the spiral arm needs to be rotated around the Galactic
Centre in the counter-clockwise direction by that amount to fit the
`magnetic arm' along which the regular component of the field reside.
A least-square fit to the data gives $\vec{B_{\rm o}}$ as 1.7$\pm$1.0 $\mu$G
and $\theta = 12^{\circ} \pm 8^{\circ}$ as shown by the solid triangles in
Fig.~\ref{fig5}. There has been several previous attempts
to estimate $B_{\rm o}$. For example Rand \& Kulkarni (1989) and Lyne \& Smith (1989) 
estimated $B_{\rm o}\sim 1.5\mu$G. More recently Han \& Qiao (1994)
and Indrani \& Deshpande (1998) estimated $B_{\rm o}\sim 1.8\mu$G.
Our value of $B_{\rm o}$ is consistent with all the previous estimates.

While the significance is not
very high, the positive value of $\theta=12^\circ \pm 8^{\circ}$ may
indeed indicate that `magnetic arm' is at some angle to the spiral
arm. In fact we can now use our model to find the pitch angle of the 
`magnetic arm'
which we define as the angle between the tangent to the arm and the
tangent to the logarithmic spiral in the direction $l=180^{\circ}$.
The model of the Perseus arm by Georgelin \& Georgelin (1976)   
gives a pitch angle of $\sim -19^{\circ}$ while our phase offset  
model considered above gives a pitch angle for the `magnetic arm' 
of $-10^{\circ}\pm8^{\circ}$, which is in good agreement with the 
earlier estimate of $-8^{\circ}$ (e.g. Han et al. 1999, Indrani \& Deshpande 1998).
\footnote{We use the negative sign for the pitch angle which is
according to the convention used by Han et al. (1999).}
However, a larger data set and an analysis extending to a larger
longitude range is obviously essential to discern this effect with
more sophisticated models than applied here as a first approach.

Indeed, the model of the regular field considered here is, almost
certainly, far too simple. It is quite likely that the magnetic field
amplitude changes as a function of pulsar distance, a situation
which was considered in detail by Rand \& Lyne (1994). A crude way to
investigate this is to study the variation of pulsar RM with
distance. For a uniform magnetic field and constant $n_e$, we expect a
linear increase of RM with distance. Assuming that no sign
reversals are present, any significant deviation would, in contrast,
point to changes in magnetic field amplitude or $n_e$. However, we are
confident that the distribution of $n_e$ along the line-of-sight to
pulsars in our selected sample is relatively uniform, so that we can
attempt this exercise for the magnetic field. 
In order to avoid problems caused by
possible uncertainties in the distance model, we study three separate
sections, i.e.~ $85^{\circ} < l < 130^{\circ}$, $130^{\circ}< l <
180^{\circ}$ and $180^{\circ}< l <245^{\circ}$, respectively. In all
cases, the data are consistent with a constant magnetic field amplitude. 
We show the RM versus DM dependence for all the $l$ range 
in Fig.~\ref{fig6}. The slope in 
the figures are proportional to the $B_{\parallel}$ but
no significant change in slope is evident. However the spread of the 
data points increases noticeably with DM, an effect which we discuss
further below.

The formal uncertainty of our determined $B_{\circ}$ value is 1$\mu$G which
is also visible from Figure~\ref{fig5}. This is much larger than the
relative error of the fitted $B_\parallel$ values which is typically
about 5 to 10\% and dominated by the uncertainties in RM. We can understand
this discrepancy by considering that our model does not take into account any
possible small-scale fluctuations of the magnetic field in strength and direction
which may be correlated with the local electron density (Beck~2001).

It is possible that many more HII regions, too faint to be detected by current surveys, 
are associated with the random component of the magnetic field. 
Sokoloff et. al. (1998) and Gaensler et al (2001) studied various effects 
and could relate the fluctuation of RM's, $\sigma_{\rm_{rm}}$, as
\begin{equation}
\sigma_{\rm_{rm}}=0.81 \times \delta n_e B_r\sqrt{L_r D}~~~\rm{rad~m^{-2}}
\label{eq2} 
\end{equation} 
where $\delta n_e$ is the fluctuation of the electron density 
in cm$^{-3}$ and $B_r$ in $\mu$G the random component of the 
magnetic field towards the pulsar. While $D$ is the distance 
to the pulsar measured in pc, $L_r$ (in pc) is the length scale 
over which $B_r$ fluctuates.  For a determination of the involved 
parameters, Rand \& Kulkarni (1989) studied pulsars with small angular 
and linear separations.  They introduced a model where they divided 
the distance to each pulsar into a number of single cells of constant 
length $L_r$ with randomly oriented field components of strength $B_r$. 
As a result, they found a length scale of $L_r\sim$50 pc and random magnetic 
field of $B_r\sim$ 5$\mu$G. This small value of $L_r$ can be compared 
to values of 100 pc and 1000 pc, respectively, which Rand \& Kulkarni 
(1989) inferred for the North Polar Spur and an area which they called 
``Region A'' where they found the magnetic field to be highly organized.

Interestingly, we can use the data obtained for our Region 1 to obtain 
another, independent estimate for the parameters. Here we can infer the 
excess in electron density $\delta n_e$ from the increased DM and 
the observed transverse length scale of about $L_r\sim$30 pc 
as mentioned in Section~\ref{sec3}. Further using the increased RM and the
estimated $\delta n_e$ and $L_r$, 
we obtain a value of $B_r\sim$ 5.7$\mu$G. Both $L_r$ and $B_r$ estimated
for Region 1 are in good agreement with the estimates of Rand \& Kulkarni (1989)
as mentioned above. 
Thus, we can demonstrate here that pulsars lying behind HII regions 
can be used as probe to find the local turbulent random 
component of the magnetic field $B_r$.

Based on our model for the regular component of the magnetic field, we can
estimate the value of $\sigma_{\rm_{rm}}$ by using the post-fit
residuals $\mid B_{\parallel}-B_{\rm model} \mid$, obtained after applying
our magnetic field model to the data (solid line in Fig.~\ref{fig5}).
These residuals are converted to residuals in RM using Eqn.~\ref{eq1}.
In Fig.~\ref{fig7} the open circles show
the residual of RM's as a function of the CL02 
model distance. To find $\sigma_{\rm_{rm}}$ we consider distance
intervals of 1 kpc. For each such bin we 
equate $\sigma_{\rm_{rm}}$ to the square root 
of the mean of the squares of the RM residuals.
The filled squares in Fig.~\ref{fig7} show the estimated
$\sigma_{\rm_{rm}}$ placed at the centre of the distance bins
considered. The estimated error bars reflects the number of points
available in each bin. We have not used PSR J2325+6316
which is at a distance of $\sim$8 kpc as there is only one pulsar
in that distance bin. 
The increase in $\sigma_{\rm_{rm}}$ with DM (or distance) seen here 
is consistent with the aforementioned increasing scatter in the RM-DM
plot of Fig.~\ref{fig6}.

The dashed line shown in Fig.~\ref{fig7} corresponds to $40\times D^{0.5}$.
We show this line only for illustration to compare the data to
a $\sqrt{D}$-dependence which is expected from Eqn.~\ref{eq2}
if the parameters
$\delta n_e$, $B_r$ and $L_r$ are distance independent. 
While the data are not adequate to claim such a $\sqrt{D}$-dependence,
we can certainly see that $\sigma_{\rm_{rm}}$ increases with distance.

\section{Field reversals revisited}
\label{frev}

Several previous studies have argued for field reversals in 
the Perseus arm and beyond based on specific behaviour of the RM 
distribution of pulsars and EG point sources (Lyne \& Smith 1989, 
Clegg et al. 1992, Han et al. 1999, 2002). A lucid description of the 
various reasons leading to conclusions regarding field reversals can 
be found in Han et al. (1999). Culmination of all these studies leads
to the proposal that a field reversal on Galactic scale is observed 
beyond the Perseus arm. While the large scale magnetic field is seen 
to follow a clockwise direction between the Perseus and the Carina-Sagittarius
arm, it is argued that the magnetic field beyond the Perseus arm
follow a counterclockwise direction.

The paucity of pulsars beyond a distance of $\sim$6 kpc forces us to rely only
on the RM's of EG sources to study the magnetic field beyond the Perseus arm.
As discussed in Brown (2002) and clarified further in Brown et al. (2003; in
preparation) one needs EG sources which are confined to a narrow latitude range.
Here we revisit the question of field reversal towards the Galactic
longitude range of $85^{\circ} <$ l $ < 240^{\circ}$ based on our findings of
anomalous RM of pulsars in Region 1 and 2.  In Fig.~\ref{fig6} we show the
behaviour of RM with DM (left panels) and RM with CL02 model distances D
(right panels) to pulsars in three specific longitude ranges.  The pulsars
affected by the HII regions and supernova remnants are indicated by open
circles in the figure. If the pulsars with open circles are excluded, then the
RM of the filled points are seen to show an overall mean increase with DM as
shown by the dashed line in the figure, with no significant change in slope
with DM (or D).  The slope is proportional to the average $B_{\parallel}$
component and no noticeable change in $B_{\parallel}$ upto a distance of 5--6
kpc is observed.  While the spread in the RM is seen to increase with both DM
and D in the shape of a cone as expected from Eqn.~\ref{eq2}.  Thus the data
seems to be consistent with a constant magnetic field upto a distance of 5--6
kpc towards the Persues arm. We can compare to this to the results by Han et
al. (1999) who suggested a large scale field reversal in our 
Region 2 at about 5--7 kpc.
The conclusion for this particular region was based on the observation that
the RMs of pulsars and EG sources beyond this distance seemed to become less negative
which can also be seen in the uppermost left panel of Fig.~\ref{fig6}.  While
their is considerable scatter in the EG sources, we have shown that RM of the
pulsars in this regions are affected by HII regions. The argument
for the suggested field reversal is hence weakened. Also for Region 1 
we have shown that the positive contribution of the RM arises from the 
HII region which further weakens the hypothesis of the reversed 
field as was suggested by Han et al. (1999).

Considering the electron density scale height to be 1 kpc, only sources that
lie within $|b| \pm 10^{\circ}$ will sample the magnetic field for a distance
larger than 6 kpc. However, not many RM measurements are currently available
for such low latitudes. In the past, Lyne \& Smith (1989) used 19 EG sources
lying within $|b| \pm 30^{\circ}$ from the Simard-Normandin \& Kronberg (1980)
catalogue and noted that the mean RM for these sources were notably smaller than
that of the distant pulsars and considered this to be indicative of field
reversal in the outer Galaxy.  However there are only 5 EG sources which lie
within $|b| \pm 10^{\circ}$, so that their conclusion and also that of Han et
al.~(1999) for Region 2 might have been influenced by inadequate statistics.

The ongoing Canadian Galactic Plane Survey
(CGPS , Landecker et al 2000) is an effort towards augmenting RM
statistics for EG sources (Brown \& Taylor 2001) which should improve our
understanding of the magnetic field in the outer Galaxy significantly.  As
mentioned earlier that their initial results already suggest RM values of
$\sim \pm$ 400 within $|b| \le 4^{\circ}$.  Further the effect of HII regions
also needs to be taken into account for such a study.

\section{Discussion and Conclusion}
\label{sec5}

In this paper we investigated the effect of HII regions on
observed pulsar RMs, isolating the region in the direction of the
Perseus arm between Galactic longitude $85^{\circ} < l < 245^{\circ}$
and Galactic latitude $|b|<10^{\circ}$ for our study. We find that
there are two regions as discussed in Section~\ref{sec4} where RM and
DM of pulsars increases anomalously due to presence of small scale HII
regions along the line-of-sight. We present a comparative study of
these regions seeking evidence from H$\alpha$ maps and further
following the nature of the emission as evident from $\lambda$21cm and 
$\lambda$11cm
radio continuum. For Region 1 there are two pulsars which are strongly
affected due to presence of the HII region S205. Based on the
pulsars' RM and DM estimates we conclude that the increased RM observed
in the centre of S205 can be well explained by an increase in 
electron density within this region. Interestingly, 
this region seems to be associated with a region where locally the
magnetic field is directed towards the observer as illustrated in
Fig.~\ref{schem}.  For Regions 2 we find 4 pulsars to be lying
behind HII complexes which explains an observed RM and DM anomaly. 
Demonstrating how the presence of HII regions strongly affects the RM of
pulsars, we emphasize the need for a carefully selected sample of pulsars 
that is unaffected by line-of-sight effects when studying the
regular large scale component of the magnetic field.

In earlier studies of the Galactic magnetic field the effect of large
supernova remnant like the North polar spur and the Gum Nebula has been taken
into account (Rand \& Kulkarni 1989, Indrani \& Deshpande 1998).  This region
spans a large angular scale in the sky and thus the pulsars behind these
regions are excluded for the analysis of large scale magnetic field supposing
that these region introduce RM anomalies.  Given the limited number of pulsars
available for any statistical analysis, however it is not entirely reasonable
to reject all pulsars behind large HII complexes.  This is also true as the
HII regions are clumpy and the pulsars might be located in regions where the
RM is relatively unaffected by the HII regions. Such information can only be
obtained by careful study of H$\alpha$ and radio continuum emission as we have
shown in this paper. A lack of such analysis might lead to several
controversial conclusions.  For demonstration, we revisited Han et
al.~(1999)'s argument that the RM's in direction of $l\sim113^{\circ}$ (our
Region 2) give indications for a field reversal on Galactic scales at a
distance of 5--7 kpc. As we have shown, this region is strongly influenced by
the presence of ionized HII complexes.  While the paucity of distant pulsars
prevents drawing firm conclusions for beyond the Perseus arm, our analysis
excludes a reversal of magnetic field towards $85^{\circ} < l < 245^{\circ}$
for distances of less than 5--6 kpc.  As mentioned before Indrani \& Deshpande
(1998) in their analysis employed a selection criteria where they excluded all
pulsars which has $B_{\parallel} > 3\mu$G.  We however note that in our sample
all the pulsars affected by HII region have $B_{\parallel} < 3\mu$G (see
Tab.~\ref{tab1}), and thus would not have been excluded in their analysis.

Further evidence of RM of extra-galactic sources being affected by HII regions
have been noted by Brown \& Taylor (2001) in the region $l=92^{\circ}$ and
$b=0.5^{\circ}$. In the recent CGPS the RM of about 380 sources are 
determined (Brown \& Taylor 2001).
All these sources lie in the $|b|<4^{\circ}$ of the Galactic plane. This is
in contrast to the data of Simard-Normandin \& Kronberg (1980) who had data on
sources away from the Galactic plane. In particular in the region $l=92^{\circ}$
and $b=0.5^{\circ}$, Brown \& Taylor (2001) notes that most of the extra-galactic sources have negative
RM's. The few pulsars with determined RM have values less than the nearby
extra-galactic sources. In three directions a sudden reversal of the
sign of RM is seen. These regions are towards diffuse continuum emission, presumably
HII regions. Based on this observation Brown \& Taylor concluded that the magnetic
field reversal correlates with nearby Galactic emission structures and not
reversals on larger scales in the Galactic spiral arms. This conclusion
is supported by our studies of Regions 1 and 2. Although
comparison with H$\alpha$ data for these anomalous regions are essential, as sometimes the thermal HII
is not visible in the continuum radio emission, either due to poor sensitivity or due
to presence of non-thermal emission. This effect is clearly seen for Region 1 and Region 2
in Fig.~\ref{fig4}.

In summary, we have presented several arguments demonstrating
that {\em HII regions strongly affect the RMs of pulsars.}
We have obtained 11 new RM of pulsars towards the Perseus arm
which gives improved statistics to study the Galactic magnetic field 
in this direction.
Using these new as well as previously published data we find 
36 pulsars out of 45 available in the region of investigation which 
are unaffected by the presence of prominent HII regions. This careful
selection helps to study the regular component of the magnetic field
$B_{\parallel}$ in the Galactic disk Section~\ref{sec4}. We derive
the following conclusions.

a) The regular large scale component of the magnetic 
field has a magnitude of $1.7\pm1.0\mu$G. 
The magnetic field follows the local Perseus arm rather well
between $150^{\circ} < l < 245^{\circ}$. There is significant
deviation observed beyond l$\sim 150^{\circ}$. 
This deviation can be
explained by considering the `magnetic arm' to be at an angle 
of 12$^{\circ}\pm8^{\circ}$ with respect to the spiral
arm or a pitch angle of $-10^{\circ}\pm8^{\circ}$. 

b) We find the fluctuation in RM to increase
as a function of distance to the pulsar. 

c) We do not find any evidence for a large-scale reversal of the 
Galactic magnetic field in direction of 
the Perseus arm at $85^{\circ} < l < 245^{\circ}$
upto a distance of 5--6 kpc.

\begin{acknowledgements}

We are grateful to the helpful comments of  the anonymous referee
which enabled us to considerably refine model of the large scale field structure.
We thank D. Backer for help in the development of the 
observing strategy and O. L\"ohmer for help during the observations.
We have benefited a lot from discussions with 
A. Shukurov, A. Fletcher, D. Sokollof, 
E. Berkhuijsen, R. Beck, S. Johnston and W. Reich. 
We thank D. Lorimer, J. A. Brown and  Graham-Smith for careful reading of 
the manuscript and their valuable comments.
We thank P. M\"uller for his help in converting the 
FITS table available
at the WHAM web site to produce the H$\alpha$ image in Region 2 as 
shown in Fig.~\ref{fig4} and also for his help in getting the VTSS 
image in the l, b coordinates. We thank P. P. Kronberg for providing us
with the unpublished RM data of extragalactic sources.
This paper uses data provided by Sharpless (1959) as distributed by 
the Astronomical Data Center at
NASA Goddard Space Flight Center. 
We would like to acknowledge the Wisconsin H-alpha mapper (WHAM) and 
Virginia Tech Spectral-Line Survey (VTSS), which is
supported by the National Science Foundation, for making available the survey for 
public use.  
The radio continuum images were obtained from the survey sampler which is 
supported by Max-Planck Institut f\"ur Radioastronomie.
We thank W. Fussh\"oller for technical help.
\end{acknowledgements}

\oddsidemargin=-0.0cm
\evensidemargin=0.5cm
\def\baselinestretch{1.0}

\begin{table*}
\caption[]{
Pulsars included in this study. We list the pulsar name, its
Galactic longitude and latitude, dispersion measure, rotation measure
and the distance and z-height as derived from the CL02
(2002) model. We also quote the magnetic field $B_{\parallel}$ as
estimated from Eqn.~\ref{eq1} (last column). Pulsars with new RM measurements
reported in this paper are highlighted in boldface.
The DMs and RMs of pulsars marked with a G are unlikely to 
be strongly affected by HII regions (see text for further details).
\label{tab1}}
\centerline{
\begin{tabular}{lrrr@{$~\pm~$}lr@{$~\pm~$}lrrr@{$~\pm~$}l}
\hline
\hline
\noalign{\medskip}
\multicolumn{1}{c}{Pulsar} &  \multicolumn{1}{c}{$l$} & 
\multicolumn{1}{c}{$b$} & 
\multicolumn{2}{c}{DM} & \multicolumn{2}{c}{RM} & 
\multicolumn{1}{c}{$D_{CL}$} & 
\multicolumn{1}{c}{$z$} & 
\multicolumn{2}{c}{$B_{\parallel}$ } \\
\multicolumn{1}{c}{J2000} & 
\multicolumn{1}{c}{(deg)} & 
\multicolumn{1}{c}{(deg)} & 
\multicolumn{2}{c}{(pc cm$^{-3}$)}& \multicolumn{2}{c}{(rad m$^{-2}$)}& 
\multicolumn{1}{c}{(kpc)} & 
\multicolumn{1}{c}{(kpc)} & 
\multicolumn{2}{c}{($\mu$G)}  \\ 
\noalign{\medskip}
\hline
\noalign{\medskip}
Region 1   &\multicolumn{9}{c}{} \\ 
0332+5434$_{\rm G}$  &  145.00 & $-$1.22 & 26.776 & 0.005 &$-$63.7 & 0.4 &  0.98 & $-$0.02&  $-$2.93 & 0.02 \\
{\bf 0343+5312}$_{\rm G}$ &147.0 & $-$1.43 & 69 & 2      &$-$84  & 20  &  2.05 & $-$0.05&  $-$1.5 & 0.5  \\
0357+5236  &  149.10 & $-$0.52 &103.650 &0.012   &261   & 13  & 2.78 & $-$0.03&   3.10 & 0.2 \\
0358+5413  &  148.19 &  0.81 & 57.14& 0.06 & 79 & 4 &  1.45 &  0.02& 1.70& 0.1 \\
\noalign{\medskip}
Region 2   &\multicolumn{9}{c}{} \\ 
{\bf 2229+6205}$_{\rm G}$    &107.15 &  3.64 &122.6&0.4 &$-$125&22 & 3.96 & 0.25 &$-$1.2&0.3  \\
2257+5909    &108.83 & $-$0.57 &151.070&0.012 &$-$322&11 &  4.50 &$-$0.04 &$-$2.63&0.1 \\
2308+5547$_{\rm G}$    &108.73 & $-$4.21 & 47.0&0.4 &$-$34&3 &  2.17 &$-$0.16 &$-$0.89& 0.1 \\
2321+6024    &112.09 & $-$0.57 & 94.78 & 0.11 &$-$230 & 10 &  3.03 &$-$0.03 &$-$2.99 & 0.1 \\
2325+6316$_{\rm G}$    &113.42 &  2.01 &195&5 &$-$102&14 & 8.08 &0.28 &$-$0.64&0.1 \\
2326+6113    &112.95 &  0.00 &122.69&0.02 &$-$221&10&  4.86 &0.0 &$-$2.22&0.1 \\
{\bf 2337+6151}    &114.28 &  0.23 & 58.38&0.09 &$-$100&18 &  3.15 & 0.01& $-$2.1&0.3 \\
\noalign{\medskip}
Others   &\multicolumn{9}{c}{} \\ 
0040+5716$_{\rm G}$    &121.45 & $-$5.57 & 90.6&0.5 &  9&13 &  2.98 &$-$0.29 & 0.12&0.2 \\
{\bf 0056+4756}$_{\rm G}$& 123.79 &$-$14.92 & 18 & 2 & $-$23 & 22&1.03  &$-$0.26&$-$1.6&1.5 \\
{\bf 0102+6537}$_{\rm G}$ &124.08 &  2.77 & 65.84&0.13 & $-$94&15 &  2.29 &0.11 & $-$1.8&0.2\\
{\bf 0108+6905}$_{\rm G}$ &124.46 &  6.28 & 59.5&0.6 & $-$46&19 &  2.22 &0.24 & $-$0.9&0.4\\
0108+6608$_{\rm G}$   &124.65 &  3.33 & 30.15&0.10 &$-$29&3 &  1.40 &0.08 &$-$1.19&0.1 \\ 
0139+5814$_{\rm G}$   &129.22 & $-$4.04 & 73.75&0.10 &$-$90&4 &  2.87 &$-$0.2 &$-$1.50&0.1 \\
0141+6009$_{\rm G}$   &129.15 & $-$2.11 & 34.80&0.10&$-$48&3 &  2.18 &$-$0.08 &$-$1.70&0.1 \\
0147+5922$_{\rm G}$   &130.06 & $-$2.72 & 40.10&0.01&$-$19&5 &  2.22 &$-$0.10 &$-$0.58&0.2 \\
0157+6212$_{\rm G}$   &130.59 &  0.33 & 29.8&0.3 &$-$29&7 &  1.68 & 0.01 &$-$1.20&0.3 \\
{\bf 0231+7026}$_{\rm G}$ &131.16 &  9.18 & 47&2 &$-$56&21 &  1.85 & 0.29& $-$1.4&0.4\\
0335+4555$_{\rm G}$     &150.35 & $-$8.04 & 47.16&0.02 &$-$41&20 &  1.64 & $-$0.23 &$-$1.07&0.5 \\
0406+6138$_{\rm G}$     &144.02 &  7.05 & 65.22&0.03&  9&3 &  2.12 & 0.26 & 0.17&0.05 \\
0454+5543$_{\rm G}$     &152.62 &  7.55 & 14.60&0.02& 10&3 &  0.67 & 0.09& 0.84&0.3 \\
0502+4654               &160.36 &  3.08 & 42.09&0.04&$-$43&6 &  1.39 &0.07 &$-$1.26&0.2 \\
0528+2200$_{\rm G}$ &183.85 & $-$6.89 & 50.877&0.001 &  $-$39.6&0.2 & 1.60 & $-$0.19 &$-$0.95&0.005 \\ 
0534+2200           &184.56 & $-$5.78 & 56.790 &0.001 & $-$42.3 &2.0 & 1.73 & $-$0.17  &$-$0.92&0.04 \\
{\bf 0538+2817}  &179.72 & $-$1.69 & 39.7&0.1 & $-$7&12 &  1.22 &$-$0.05 & $-$0.03&0.4 \\
0543+2329$_{\rm G}$ &184.36 & $-$3.32 & 77.690 &0.001 &   8.7 &0.7 & 2.06 & $-$0.12  &0.13&0.05 \\
{\bf 0612+3721}$_{\rm G}$  &175.45 &  9.09 & 26.7&0.1 & 12&20 &  0.85 & 0.13& 0.6&0.9 \\
0614+2229$_{\rm G}$ &188.79 & 2.39  & 96.70 &0.05  & 67.0  &0.7 & 2.08 &  0.08  &1.2&0.06 \\
0624$-$0424$_{\rm G}$ &213.79 &$-$8.04  & 72.0  & 1.5  & 42 & 7.0 & 2.83 & $-$0.39 &0.7&0.1 \\
0629+2415$_{\rm G}$ &188.82 & 6.22  & 84.20  & 0.03 & 82 & 4 & 2.23 &  0.24 &1.2&0.05 \\
0659+1414$_{\rm G}$ & 201.11 & 8.26   & 14.02&0.05 & 22 & 5 & 0.67 & 0.09  &1.9&0.4 \\
0729$-$1836$_{\rm G}$&233.76 & $-$.34  & 61.30 & 0.04 & 53 & 6 & 2.90 & $-$0.02 &1.1&0.1 \\
0742$-$2822$_{\rm G}$& 243.77 &  $-$2.44 & 73.77&0.02& 150.43&0.05 & 2.07 & $-$0.09 & 2.5&0.01\\
0758$-$1528$_{\rm G}$&234.46 & 7.22  & 63.7  & 0.3  & 55 & 7 & 2.96 & 0.37  &1.1&0.1 \\
2108+4441$_{\rm G}$ & 86.91 & $-$2.01  & 139.88 &0.03 & $-$14 & 9&4.96 & $-$0.17 &$-$0.1&0.1 \\
2113+4644$_{\rm G}$& 89.0    &$-$1.27   &141.50 &0.04 & $-$224 &2&4.53 & $-$0.10  & $-$1.9&0.01 \\
2149+6329$_{\rm G}$     &104.25 &  7.41 &128&1 &$-$160&7 & 5.50 & 0.71&$-$1.54&0.1\\
{\bf 2150+5247}$_{\rm G}$  & 97.52 & $-$0.92 &148.94&0.02 &$-$44&11 &  4.62 &$-$0.07 & $-$0.4&0.1 \\
2219+4754$_{\rm G}$    & 98.38 & $-$7.60 & 43.52&0.02 &$-$35.3&1.8 &  2.22 &$-$0.29 &$-$1.0&0.1 \\
2225+6535$_{\rm G}$    &108.64 &  6.85 & 36.16&0.05 &$-$21&3 &  1.86 &0.22 &$-$0.72&0.1 \\
{\bf 2242+6950}$_{\rm G}$   &112.22 &  9.69 & 40.7&0.8 &  $-$30&30 &  1.96 & 0.33 &$-$0.9&0.9 \\
2354+6155$_{\rm G}$     &116.24 & $-$0.19 & 94.34&0.06 &$-$77&6 &  3.41 & $-$0.01&$-$1.01&0.1 \\
\hline
\end{tabular}
}
\end{table*}


\begin{thebibliography}{}

\bibitem[Anderson et al 1996]{and96}Anderson, S. B., Cadwell, B. J., Jacoby, B. A. et al., 1996, ApJL, 468, L55.
\bibitem[Backer et al 1997]{b97}Backer, D. C., Dexter, M. R., Zepka, A. et al., 1997, PASP, 109, 61.
\bibitem[Beck 2001]{b2k1}Beck R., 2001, Space Sci. Rev., 99, 243.
\bibitem[Blitz, Fich and Stark 1982]{fbs82} Blitz, L., Fich, M. \& Stark, A. A., 1982, ApJS, 49, 183.
\bibitem[Broten et al 1988]{bnv88} Broten, N. W., MacLeod, J. M. \& Vallee, J. P., 1988, Ap\&SS, 141, 303.
\bibitem[Cordes \& Lazio 2002]{cl2k2} Cordes, J. M. \& Lazio, T. J. W., 2002, astro-ph/0207156
\bibitem[Jo Ann Brown 2001]{jthesis}Brown, J.C., 2002, Ph.D. thesis, University of Calgary
\bibitem[Jo Ann Brown 2001]{j2001} Brown. J \& Taylor, A. R., 2001, ApJ, 563, L31.
\bibitem[Cooper and Price 1962]{cp62} Cooper B. F. C. \& Price. R. M., 1962, Nature, 195, 1084.
\bibitem[Downs et al 1981]{dsp81}Downes, A. J. B., Salter, C. J. \& Pauls, T., 1981, A\&A, 103, 277.
\bibitem[Fesen Blair Krishner 1985]{fb85} Fesen, R. A., Blair, W. P. \& Kirshner, R. P., 1985, ApJ, 292, 29.
\bibitem[Fich and Blitz 1984]{fb84} Fich, M. \& Blitz, L., ApJ, 279, 125.
\bibitem[Fuerst et al 1984]{f1984} F\"{u}rst, E., Reich, W. \& Steube, R., 1984, A\&A, 133,11.
\bibitem[Fuerst et al 1993]{frs93}F\"{u}rst, E., Reich, W. \& Seiradakis, J. H., 1993, A\&A, 276, 470. 
\bibitem[Fuerst et al 1990]{f1990}F\"{u}rst, E., Reich, W. Reich, P. \& Reif, K., 1990, A\&AS, 85, 691.
\bibitem[Frick et al 2001]{fsss2k01}Frick, P., Stepanov, R., Shukurov, A. \& Sokoloff, D., 2001, MNRAS, 325, 649.
\bibitem[Gaensler et al 2001]{g2001} Gaensler, B. M., Dickey, John M., McClure-Griffiths, N. M. et al., 2001, ApJ, 549, 959.
\bibitem[Georgelin and Georgelin 1976]{gg76} Georgelin, Y. M. \& Georgelin, Y. P., 1976, A\&A, 49, 57.
\bibitem[Gray et al, 1998]{gldt98}Gray, A. D.; Landecker, T. L., Dewdney, P. E. \& Taylor, A. R., 1998, Nature, 393, 660.
\bibitem[Greqing \& Walmsley 1971]{gw71} Grewing, M. \&  Walmsley, M., 1971, A\&A, 11, 65.
\bibitem[Haffner et al 2001]{h2k1}Haffner, L. M., Reynolds, R. J., Madsen et al., 2001, AAS, 199, 58.01
\bibitem[Han \& Qiao]{hq94}Han, J. L. \& Qiao, G. J., 1994, A\&A, 288, 759.
\bibitem[Han et al1999]{h99}Han, J. L., Manchester, R. N. \& Qiao, G. J., 1999, MNRAS, 306, 371.
\bibitem[Han et al 2002]{h2k2}Han, J. L., Manchester, R. N., Lyne, A. G. \& Qiao, G. J., 2002, ApJ, 570, L17.
\bibitem[Hewish et. al. 1968]{hbpsc68} Hewish, A., Bell, S. J., Pilkington, J. D. H. et al., 1968, Nature, 217, 709.
\bibitem[Indrani and Deshpande 1998]{id98} Indrani, C. \& Deshpande, A. A., 1998, New Astronomy, 4, 33.
\bibitem[kim et al 1988]{kkl88}Kim, K.-T., Kronberg, P. P. \& Landecker, T. L., 1988, ApJ, 96, 704.
\bibitem[Kulkarni et al 1993]{kpha93} Kulkarni, S. R., Predehl, P., Hasinger, G.\& Aschenbach, B., 1993, Nature, 362, 135. 
\bibitem[Landekar et al 2000]{ldb2k}Landecker, T. L., Dewdney, P. E. \& Burgess, T. A., 2000,
A\&A, 145, 509.
\bibitem[Leahy \& Roger 1991]{lr91} Leahy, D. A.; Roger, R. S., 1991, AJ, 101, 1033L.
\bibitem[loehmer et al 2001]{lkddl} L\"{o}hmer, O., Kramer, M., Mitra, D, Lorimer, D. R.
\& Lyne, A. G., 2001, ApJ, 562, L157.
\bibitem[Lynds 1965]{l65} Lynds, Beverly T., 1965, ApJS, 12, 163.
\bibitem[Manchester \& Taylor]{mt77} Manchester, R. N. \& Taylor, J., 1977, San Francisco : W. H. Freeman
\bibitem[Milne 1979]{m79} Milne, D. K., AuJPh, 32, 83.
\bibitem[Mitra and Ramachandran 2001]{mr2k01}Mitra, D \& Ramachandran, R, 2001, A\&A, 370, 586.
\bibitem[muller et al 2002]{mmb2k1} M\"{u}ller, P., Mitra, D. \& 
Berkhuijsen, E., 2002, in preparation.
\bibitem[Prentice 1969]{pt69} Prentice, A. J. R. \& Ter Haar, D., 1969, MNRAS, 146, 423.
\bibitem[Rand and Kulkarni 1989]{rk89} Rand, R. J. \& Kulkarni, S. R., ApJ, 343, 760.
\bibitem[Rand and Lyne 94]{rl94}Rand, R. J. \& Lyne, A. G., 1994, MNRAS, 268, 497.
\bibitem[Reich et al 1984]{rfsr84}Reich, W., F\"{u}rst. E., Steffen, P., Reif, K.\& Haslam, C.G.T.,1984, A\&AS, 58, 197.
\bibitem[Reich et al 1990]{rrf90}Reich, W., Reich, P.\& F\"{u}rst, E., 1990, A\&AS, 83, 539.
\bibitem[Sharpless 1959]{s59}Sharpless, Stewart, 1959, ApJS, 4, 257.
\bibitem[Simonetti et al 1996]{sim96} Simonetti, J. H., Dennison, B. \& Topsana, G. A., 1996, ApJ, 458, L1.
\bibitem[Smith 1968]{s68} Smith, F. G., 1968, Nature, 218, 325.
\bibitem[simard kronberg 1980]{sk80} Simard-Normandin, M. \& Kronberg, P. P., 1980, ApJ, 242, 74.
\bibitem[Sokoloff et al 1998]{sbsbbp98} Sokoloff, D. D., Bykov, A. A., Shukurov, A. et al., 1998, MNRAS, 299, 189.
\bibitem[Taylor \& Cordes 1993]{tc93} Taylor, J. H. \& Cordes, J. M., 1993, ApJ, 411, 674.
\bibitem[Taylor, Manchester \& Lyne 1993]{tml93} Taylor, J. H., Manchester, R. N. \& Lyne, A. G., 1993, ApJS, 88, 529.
\bibitem[Uyaniker et al. 2001]{ukb2k1}Uyan{\i}ker, B., Kothes, R. \& Brunt, C. M., 2002, ApJ, 565, 1022.
\bibitem[Vall\'ee \& Bignell 1983]{vb83} Vall\'ee, J. P. \& Bignell, R. C., 1983, ApJ, 272, 131.
\bibitem[Zdanavicius et al 2001]{zzs2001}Zdanavicius, J., Cernis, K., Zdanavicius, K. \& Straizys, V., 2001, Baltic Astronomy, 10, 349.
\end{thebibliography}
\end{document}